\documentclass{IEEEtran}
\usepackage[utf8]{inputenc}
\usepackage{cite}
\usepackage{amsmath,amssymb,amsfonts}
\usepackage{algorithmic}
\usepackage{graphicx,color}
\usepackage{textcomp}
\usepackage{bm}
\usepackage{tabularx}
\usepackage{mathtools}
\usepackage{multirow}
\usepackage{subcaption}
\usepackage{flushend}

\newcommand{\Ecal}{\mathcal{E}}
\newcommand{\Tcal}{\mathcal{T}}
\newcommand{\Vcal}{\mathcal{V}}

\newcommand{\ulcapacity}{\bar{c}_{\text{UL}}}
\newcommand{\islcapacity}{\bar{c}_{\text{ISL}}}
\newcommand{\flcapacity}{\bar{c}_{\text{FL}}}

\newcommand{\Vcalut}{\mathcal{V}_{\text{U}}}
\newcommand{\Vcalsat}{\mathcal{V}_{\text{S}}}
\newcommand{\Vcalgw}{\mathcal{V}_{\text{G}}}

\newcommand{\Ecalul}{\mathcal{E}_{\text{UL}}}
\newcommand{\Ecalisl}{\mathcal{E}_{\text{ISL}}}
\newcommand{\Ecalfl}{\mathcal{E}_{\text{FL}}}
\newcommand{\Ecaltl}{\mathcal{E}_{\text{TL}}}

\newcommand{\Ecalbar}{\bar{\mathcal{E}}}

\newcommand{\nn}{\nonumber}

\title{Optimizing Satellite Network Infrastructure: A Joint Approach to Gateway Placement and Routing}

\author{
\IEEEauthorblockN{Yuma Abe\IEEEauthorrefmark{1}\IEEEauthorrefmark{2}, Flor Ortiz\IEEEauthorrefmark{1}, Eva Lagunas\IEEEauthorrefmark{1}, Victor Monzon Baeza\IEEEauthorrefmark{1}, Symeon Chatzinotas\IEEEauthorrefmark{1}, and Hiroyuki Tsuji\IEEEauthorrefmark{2}} \\
\IEEEauthorblockA{\IEEEauthorrefmark{1}Interdisciplinary Centre for Security, Reliability and Trust (SnT), University of Luxembourg, Luxembourg, \emph{\{yuma.abe, flor.ortiz, eva.lagunas, victor.monzon, symeon.chatzinotas\}@uni.lu}} \\
\IEEEauthorblockA{\IEEEauthorrefmark{2}Space Communication Systems Laboratory, Wireless Networks Research Center, Network Research Institute, National Institute of Information and Communications Technology (NICT), Japan, \emph{\{yuma.abe, tsuji\}@nict.go.jp}}
}

\begin{document}

\maketitle
\thispagestyle{empty}

\begin{abstract}
Satellite constellation systems are becoming more attractive to provide communication services worldwide, especially in areas without network connectivity.
While optimizing satellite gateway placement is crucial for operators to minimize deployment and operating costs, reducing the number of gateways may require more inter-satellite link hops to reach the ground network, thereby increasing latency.
Therefore, it is of significant importance to develop a framework that optimizes gateway placement, dynamic routing, and flow management in inter-satellite links to enhance network performance.
To this end, we model an optimization problem as a mixed-integer problem with a cost function combining the number of gateways, flow allocation, and traffic latency, allowing satellite operators to set priorities based on their policies.
Our simulation results indicate that the proposed approach effectively reduces the number of active gateways by selecting their most appropriate locations while balancing the trade-off between the number of gateways and traffic latency.
Furthermore, we demonstrate the impact of different weights in the cost function on performance through comparative analysis.
\end{abstract}

\begin{IEEEkeywords}
Satellite constellations, gateway placement, path routing, flow management
\end{IEEEkeywords}

\section{Introduction}

Satellite constellation systems, such as Starlink, OneWeb, Amazon Kuiper, and Telesat Lightspeed, are becoming increasingly important for providing communication services anywhere and anytime over the globe~\cite{Al-Hraishawi_CST23}.
To ensure broad coverage and low latency on Earth, these systems require the deployment of hundreds or even thousands of satellites in low Earth orbit.
Consequently, this necessitates the installation of many gateway stations to control the fleet and manage the traffic over these multiple satellites.

On the other hand, satellite operators are seeking to reduce the number of gateways in order to minimize deployment and operational costs.
However, reducing the number of gateways is expected to decrease the total system capacity and increase user latency~\cite{Guo_TVT21}.
Thus, there is a trade-off between the number of gateways and network performance, and satellite operators need to optimize their system configuration according to their priorities.

Previous studies on gateway placement have been roughly divided into two categories: i) in the first category, researchers aimed to identify precise gateway locations by considering multiple candidate locations and selecting a subset from them.
This approach is exemplified in~\cite{Chen_IOTJ22, Portillo_IEEEAeroConf18, Guo_TVT21}.
ii) In the second category, the focus was on dividing a defined region into a grid and then selecting specific grid segments.
Researchers investigated this method in~\cite{Chen_IJSCN21, Zhu_VTC20}.
A common approach in these studies is solving an integer problem to identify the optimal gateway locations.
In the first approach, the authors of~\cite{Chen_IOTJ22} developed an analytical hop-count model within the satellite constellation to simplify routing calculations.
They optimized gateway placement and the number of inter-satellite hops using particle swarm optimization (PSO).
In another study,~\cite{Portillo_IEEEAeroConf18}, the authors examined 77 candidate locations and used a genetic algorithm (GA) to find the optimal subset that maximizes system performance in terms of data rate.
These studies focus on identifying the best number of gateways by setting and varying a maximum gateway constraint.
Conversely, the research in\cite{Guo_TVT21} aimed to minimize the number of gateways from 471 candidates by utilizing a detailed satellite coverage model and traffic model solved by PSO.
Regarding the second approach, in~\cite{Chen_IJSCN21}, the authors divided the world into 216 grids and sought the best 30 locations.
They defined a cost function that included average time delay, maximum node load, and gateway load balance, and solved it using GA.
In this study, each gateway was assumed to be at the center of its cell without determining an exact location.
However, few studies have considered the routing algorithm within the satellite constellation network.
In~\cite{Chen_IJSCN21} and~\cite{Zhu_VTC20}, the focus was on minimizing hop count and balancing traffic load under the assumption of unlimited inter-satellite link capacity.

In this paper, we propose a joint framework for optimizing the gateway placement, routing, and flow management in satellite paths, considering a global geographical traffic distribution.
Our main contributions are: i) modeling an optimization framework to jointly determine the optimal static gateway placement and dynamic routing strategies; ii) incorporating the capacity constraints of each link, including those in inter-satellite links, into our model; and iii) integrating a weighted cost function for key performance indicators, allowing satellite operators to prioritize them based on their specific requirements.
We first introduce the system model of satellite and ground networks.
Then, we formulate an optimization problem to derive the best gateway placement, routing path, and flow allocation jointly.
In our problem setting, we assume that candidate gateway locations are obtained in advance because it is impractical to consider all locations on Earth\footnote{Usually, satellite operators deploy gateways where they already have a point of presence and where they can establish an optical ground connection to the network core, e.g., https://www.ses.com/our-coverage/teleport-map}.
To this end, we utilize the gateway geographical planning method to obtain the candidates, e.g., in~\cite{Baeza_IEEENetwork23}.
Finally, we consider the candidate gateway locations as input of our algorithm and evaluate its performance and the effect of weights in a cost function, which represents the priority for the satellite operator.

\section{System Model}

In this section, we describe a system model comprising both space and ground segments.
The symbols $t$ and $T$ represent the discrete-time step and the total time duration of the targeted finite-time horizon, respectively.
We define a set of the time as $\Tcal = \{1, 2, \ldots, T\}$.
In the following modeling phase, we will omit the discrete time index $t\in\Tcal$ in our notation for the sake of simplicity.

\subsection{Satellite and Ground Network Configuration}

\begin{figure}[tb]
    \centering
    \includegraphics[width=0.9\columnwidth]{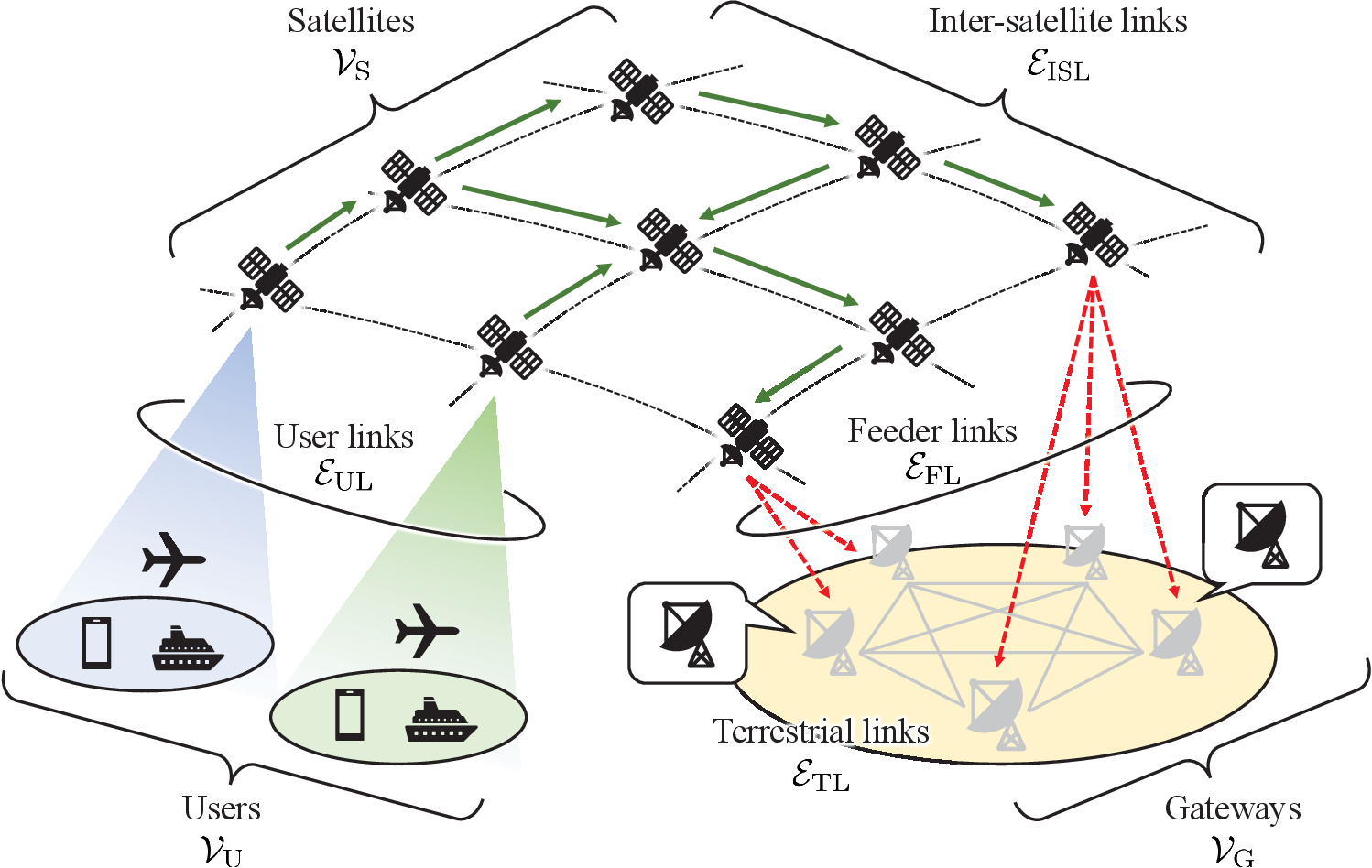}
    \caption{Overview of the joint optimization for gateway placement and routing. Gateways colored in grey are candidates of this study, and the problem is finding the best selection from these candidates considering the appropriate low-latency routing strategy.}
\label{fig:overview}
\end{figure}

This paper focuses on a network that integrates both satellite and ground components, encompassing users, satellites, and gateways, as shown in Fig.~\ref{fig:overview}.
Users, such as aircraft and ships, request communication links with satellites, which then function as routing nodes to provide the links for the users.
Gateways, situated in the ground segment, facilitate connections between users, satellites, and the terrestrial network.

We model this network as a graph, denoted by $(\Vcal, \Ecal)$, where $\Vcal$ and $\Ecal$ represent the sets of all nodes and links, respectively.
Subsets $\Vcalut, \Vcalsat$, and $\Vcalgw$ correspond to sets of users, satellites, and gateways, respectively; hence, $\Vcal = \Vcalut \cup \Vcalsat \cup \Vcalgw$.
Additionally, $N_{u} = |\Vcalut|$, $N_{s} = |\Vcalsat|$, and $N_{g} = |\Vcalgw|$ denote the numbers of users, satellites, and gateways, respectively.
It is important to note that the number of gateways refers to potential candidates, and the actual number of operational gateways will be determined later.

We identify four types of links: i) user link, between a user and a satellite; ii) inter-satellite link, between satellites; iii) feeder link, between a satellite and a gateway; and iv) terrestrial link, between gateways.
The respective sets of these links are $\Ecalul, \Ecalisl, \Ecalfl$, and $\Ecaltl$, with the entire set of links represented by $\Ecal = \Ecalul \cup \Ecalisl \cup \Ecalfl \cup \Ecaltl$.
The viability of these links is based on factors like relative positioning and link status, which are subject to change due to satellite movements, weather conditions, traffic congestion, and potential failures.
Consequently, the set of potential links varies over time.
For the scope of this paper, we concentrate solely on traffic flowing from users to gateways.
Accordingly, user and feeder links are directional from user to satellite and from satellite to gateway, respectively, while inter-satellite and terrestrial links are bidirectional.

The propagation latency of traffic is primarily influenced by the distance between two nodes and the speed of the signal.
Our focus is on distance-based latency, excluding processing delays in satellites or gateways.
In user, inter-satellite, and feeder links, signal transmission occurs at the speed of light in a vacuum, represented by $c$.
On the other hand, for terrestrial links, the signal speed is assumed to be $2c/3$, accounting for the refractive index of 1.5 in optical fibers.

Finally, we introduce two functions, $u(e)$ and $v(e)$, to map an edge $e$ (i.e., communication link) to its starting and ending nodes, respectively.
These functions are instrumental in representing both the nodes and their indices.

\subsection{Constellation Model}

Assuming the constellation consists of $P$ orbital planes, with each plane containing $K$ satellites, the total number of satellites in the network amounts to $N_{s} = PK$.
We describe these satellites as forming a full-grid mesh network.
This means that each satellite is connected to two other satellites within the same orbit (i.e., intra-orbit links) and to two satellites in adjacent orbits (i.e., inter-orbit links).
The former is always maintained, while the latter continually changes depending on the situation.
Consequently, each satellite is equipped with a minimum of four inter-satellite link terminals to facilitate these connections.
Additionally, there are two more terminals dedicated to user and feeder links.

\subsection{Gateway Model}
\label{sec:gateway_model}

We assume a predefined set of gateway candidates, as outlined in~\cite{Baeza_IEEENetwork23}.
Additionally, it is presumed that all these gateways are interconnected through terrestrial optical links, forming a full-grid mesh topology.
This arrangement implies that once traffic reaches any gateway, it can be reliably routed to the intended destination gateway, ensuring connectivity.

We also assume that each gateway has a specific infrastructure or server and handles a particular service.
Because of this, each traffic has a specific destination gateway.
This will be further explained in the following subsection.

\begin{table*}[tb]
\centering
\caption{A list of decision variables and parameters in this paper.}
    \scalebox{1}{
        \begin{tabular}{cl} \hline
            Decision variables & Description \\ \hline
            $x_{\ell}$ & The binary decision variable of gateway placement for $\ell$-th gateway \\
            $b_{i}$ & The continuous decision variable of flow allocated to $i$-th user \\
            $y_{i,e}$ & The binary decision variable of a link $e$ is assigned to $i$-th traffic \\
            $f_{i,e}$ & The continuous decision variable of a flow in a link $e$ for $i$-th traffic \\
            $z_{e}$ & The binary decision variable indicating whether nodes $u(e)$ and $v(e)$ are connected in a feeder link $e$ \\
            \hline
            Parameters & Description \\ \hline
            $N_{u}, N_{s}, N_{g}$ & Number of users, satellites, and gateways \\
            $\Vcal,\Ecal$ & Network graph, sets of all nodes and links in the network \\
            $\Vcalut, \Vcalsat, \Vcalgw$ & Sets of users, satellites, and gateways \\
            $\Ecalul, \Ecalisl, \Ecalfl, \Ecaltl$ & Sets of user links, inter-satellite links, feeder links, and terrestrial links \\
            $\Ecalbar_{i}$ & A set of links assigned to $i$-th traffic \\
            $u(e), v(e)$ & Starting and ending nodes of an edge $e$ \\
            $\ulcapacity, \islcapacity, \flcapacity$ & Capacity of user links, inter-satellite links, and feeder links \\
            $r_{i}$ & Flow requirement of $i$-th traffic \\
            $G_{i}$ & Destination gateway of $i$-th traffic \\
            $\ell_{i}$ & Latency of $i$-th traffic to reach the destination gateway \\
            $\bar{\ell}_{e}$ & Propagation latency of a link $e$ \\
            $s_{i}$ & Normalized flow gap ratio of $i$-th traffic \\ 
            \hline
        \end{tabular}
    }
\label{table:variables_parameters}
\end{table*}

\subsection{User and Traffic Model}

We assume the presence of $N_{u}$ user-generated traffic flows.
Each traffic is characterized by a set of requirement parameters: the amount of flow requirement $r$ and destination gateway $G$.
These parameters are denoted by a tuple $(r_{i}, G_{i})$ for the $i$-th traffic flow.
Traffic originates from users and is routed towards the destination gateway, potentially passing through other gateways.
We refer to a gateway that connects to a satellite via a feeder link and routes traffic to the final destination gateway as a ``relay gateway'' in this paper.
When traffic uses a relay gateway, the latency related between the relay and destination gateways contributes to the overall network performance.

It is important to note, as specified in Section~\ref{sec:gateway_model}, that the gateways discussed here are merely candidates.
Consequently, the designated destination gateway for a traffic flow might not be physically installed.
To address this, as already mentioned in Section~\ref{sec:gateway_model}, we assume that the destination gateway for each traffic flow corresponds to a specific infrastructure or server located near the gateway candidate.

\section{Problem Formulation}

In this section, we formulate an optimization problem aimed at achieving a dual objective: minimizing the number of gateways and maximizing user satisfaction.
User satisfaction is quantified based on compliance with the flow requirement and the latency to reach the destination gateway, which are critical factors in the performance of the network.
The decision variables and parameters used in this formulation are comprehensively summarized in Table~\ref{table:variables_parameters}.

\subsection{Decision Variables}

We begin by defining the decision variables for the gateway placement problem.
Firstly, a binary variable representing gateway placement is defined as follows:
\begin{align}
    x_{\ell} =
    \begin{cases}
        1, & \text{if $\ell$-th gateway is placed,} \\
        0, & \text{otherwise.}
    \end{cases}~~\forall \ell \in \Vcalgw
\label{eq:gateway_binary}
\end{align}

Next, we describe the flow allocation for each traffic.
The amount of flow intended for the $i$-th traffic is denoted by $b_{i} \geq 0$.
Additionally, we define a binary variable for link assignment, $y_{i,e}, \forall i \in \Vcalut, e \in \Ecal$, as follows:
\begin{align}
    y_{i,e} =
    \begin{cases}
        1, & \text{if $i$-th traffic is assigned to a link $e$,} \\
        0, & \text{otherwise.}
    \end{cases}
\label{eq:link_binary}
\end{align}
Here, the set of links assigned to the $i$-th traffic is represented by $\Ecalbar_{i} = \{e \mid y_{i,e} = 1\}$.
We denote $f_{i,e} \geq 0, \forall i \in \Vcalut, e \in \Ecal$, as the flow amount on link $e$ for the $i$-th traffic.


Finally, to describe constraints on feeder link connections, which will be discussed in Section~\ref{sec:constraints}, we introduce a binary variable, $z_{e}, e \in \Ecalfl,$ defined by
\begin{align}
    z_{e} =
    \begin{cases}
        1, & \text{if a link $e$ is actually used as a feeder link,} \\
        0, & \text{otherwise.}
    \end{cases}
\label{eq:link_binary_z}
\end{align}

\subsection{Constraints}
\label{sec:constraints}

We define constraints for the optimization problem of gateway placement and routing, such as feasibility constraints, connection constraints, capacity constraints, and flow constraints.

To ensure that only feeder links associated with the deployed gateway are active, the following inequality must hold:
\begin{align}
    y_{i,e} \leq x_{v(e)},~\forall i \in \Vcalut, e \in \Ecalfl.
\label{eq:constraint_feasibility_y_x}
\end{align}
Because of this constraint, $y_{i,e} = 0$ holds when a targeted 
gateway is not active, i.e., $x_{v(e)} = 0$.
Additionally, we impose the following inequality constraint to represent the relationship between the decision variable $y_{i,e}$ and its corresponding $z_{e}$:
\begin{align}
    y_{i,e} \leq z_{e},~\forall i \in \Vcalut, e \in \Ecalfl.
\label{eq:constraint_feasibility_y_z}
\end{align}
This constraint has the same function as Eq.~(\ref{eq:constraint_feasibility_y_x}) regarding $z_{e}$.

To guarantee that traffic flows only through active links, the following inequality is required:
\begin{align}
    f_{i,e} \leq M y_{i,e},~\forall e \in \Ecal,
\label{eq:constraint_feasibility_f_y}
\end{align}
where $M$ is a constant, which is called as Big-M~\cite{Bazaraa_09}\footnote{This constant, $M$, is used to express the relationship between a binary variable and a real variable when we want the real variable to have a value when the binary variable is 1. If $M$ is set smaller than the possible value of the real variable, then this inequality constrains the real variable. Therefore, $M$ must be set sufficiently large. However, setting $M$ too large may result in instability when solving the optimization problem. In the following simulation, $M$ is set equal to the capacity of inter-satellite links.}, and is chosen as large enough compared to $f_{i,e}$.

In this paper, we assume that users can connect to at most one satellite in the user links, and similarly, satellites and gateways can connect with at most one gateway and one satellite, respectively, in feeder links. 
These constraints are expressed as:
\begin{subequations}
\label{eq:constraint_connection_num}
\begin{align}
    &\sum_{e\in\Ecalul} y_{i,e}  \leq 1,~\forall i \in \Vcalut, 
    \label{eq:constraint_connection_num_user} \\
    &\sum_{j=u(e), e \in \Ecalfl} z_{e} \leq 1,~\forall j \in \Vcalsat,
    \label{eq:constraint_connection_num_satellite} \\
    &\sum_{\ell=v(e), e \in \Ecalfl} z_{e} \leq 1,~\forall \ell \in \Vcalgw,
    \label{eq:constraint_connection_num_gateway}
\end{align}
\end{subequations}
respectively.

The capacity constraints for user, inter-satellite, and feeder links are as follows:
\begin{subequations}
\label{eq:constraint_capacity}
\begin{align}
    &\sum_{j=v(e), e\in\Ecalul} f_{i,e} \leq \ulcapacity,~\forall j \in \Vcalsat,
    \label{eq:constraint_capacity_ul} \\
    &\sum_{\substack{i \in \Vcalut}} f_{i,e} \leq \islcapacity,~\forall e \in \Ecalisl,
    \label{eq:constraint_capacity_isl} \\
    &\sum_{i \in \Vcalut} f_{i,e} \leq \flcapacity,~\forall e \in \Ecalfl,
    \label{eq:constraint_capacity_fl}
\end{align}
\end{subequations}
where $\ulcapacity, \islcapacity$, and $\flcapacity$ are the maximum capacities of user, inter-satellite, and feeder links, respectively.
In this study, these capacities are assumed to be uniform for each type of link.
The capacity of terrestrial links is considered unlimited.

To ensure efficient resource utilization, the network is designed to maintain the balance of in-flow and out-flow for each link and traffic.
Thus, the following conservation laws are required:
\begin{subequations}
\label{eq:constraint_flow}
\begin{align}
    &b_{i} = \sum_{e\in\Ecalul} f_{i,e},~\forall i \in \Vcalut, 
    \label{eq:constraint_flow_user} \\
    &\sum_{\substack{j = v(e), \\ e\in\Ecalul}} f_{i,e} + \sum_{\substack{j = v(e), \\ e\in\Ecalisl}} f_{i,e} = \sum_{\substack{j = u(e), \\ e\in\Ecalisl}} f_{i,e} + \sum_{\substack{j = u(e), \\ e\in\Ecalfl}} f_{i,e},
    \label{eq:constraint_flow_sat} \\
    &\sum_{\substack{j = v(e), \\ e\in\Ecalul}} y_{i,e} + \sum_{\substack{j = v(e), \\ e\in\Ecalisl}} y_{i,e} = \sum_{\substack{j = u(e), \\ e\in\Ecalisl}} y_{i,e} + \sum_{\substack{j = u(e), \\ e\in\Ecalfl}} y_{i,e},
    \label{eq:constraint_flow_total_num} \\
    &\forall i \in \Vcalut, j \in \Vcalsat. \nn
\end{align}
\end{subequations}
These constraints ensure the flow conservation law for the user and inter-satellite links, i.e., the flow coming into any intermediate node needs to equal the flow going out of that node.

Finally, we require constraints to ensure that each traffic flow reaches its destination gateway.
Traffic can reach the destination gateway in two ways: directly to the destination gateway in a feeder link or through a relay gateway in a terrestrial link.
For direct traffic to the destination gateway, the following constraints must be satisfied:
\begin{subequations}
\label{eq:constraint_flow_gateway_destination}
\begin{align}
    &b_{i} = \sum_{\substack{G_{i} = v(e), \\ e\in\Ecalfl}} f_{i,e} + \sum_{\substack{G_{i} = v(e), \\ e\in\Ecaltl}} f_{i,e},~\forall i \in \Vcalut, 
    \label{eq:constraint_flow_gateway_destination_flow} \\
    &\sum_{\substack{G_{i} = v(e) \\ e\in\Ecalfl}} y_{i,e} + \sum_{\substack{G_{i} = v(e) \\ e\in\Ecaltl}} y_{i,e} \leq 1.
    \label{eq:constraint_flow_gateway_destination_num}
\end{align}
\end{subequations}
The first equation establishes that each traffic flow is equal to the sum of the in-flows from both feeder and terrestrial links.
The second equation specifies that either feeder links or terrestrial links are used exclusively.

For all other gateways that are not specified as the destination gateway and function as relay gateways, the following constraints are required:
\begin{subequations}
\label{eq:constraint_flow_gateway_others}
\begin{align}
    &\sum_{\substack{\ell = v(e), \\ e\in\Ecalfl}} f_{i,e} = \sum_{\substack{\ell = u(e), \\ e\in\Ecaltl}} f_{i,e},~
    \sum_{\substack{\ell = v(e), \\ e\in\Ecaltl}} f_{i,e} = 0,
    \label{eq:constraint_flow_gateway_others_flow} \\
    &\sum_{\substack{\ell = v(e), \\ e\in\Ecalfl}} y_{i,e} = \sum_{\substack{\ell = u(e), \\ e\in\Ecaltl}} y_{i,e},~\forall i \in \Vcalut,~\ell \in \Vcalgw\setminus\{G_{i}\}.
    \label{eq:constraint_flow_gateway_others_num}
\end{align}
\end{subequations}
The first and third equations indicate that both flows and the number of links are balanced in the gateway with respect to in-flows from feeder links and out-flows to terrestrial links.
Meanwhile, the second equation specifies that there is no in-flow to the gateway on terrestrial links.

\subsection{Cost Function}

In this study, we consider the number of gateways and the network's performance, i.e., allocated flow and latency, as key metrics to be included in a cost function.
Here, allocated flow is a continuous variable, while latency is a discrete variable accounted for by network hops.
In this section, we add the time index as a subscript of each term because a cost function includes the summation over time.

We first define the normalized number of selected gateways as:
\begin{align}
    J_{g} = \dfrac{1}{N_{g}} \sum_{\ell\in\Vcalgw} x_{\ell}.
    \label{eq:cost_gateway_number}
\end{align}
This value is time-invariant and represents the number of active gateways normalized by the number of gateway candidates.

Satellite operators try to allocate as much flow as possible to meet users' requests.
To quantify this, we define the normalized flow gap ratio as the difference between the flow allocation and the flow requirement for each traffic divided by the flow requirement, as follows:
\begin{align}
    s_{i,t} = \max \left(0, \dfrac{r_{i} - b_{i,t}}{r_{i}} \right).
    \label{eq:satisfaction_flow}
\end{align}
This metric illustrates that the flow gap ratio decreases as the allocated flow for each user approaches their requirement, and it becomes zero once the allocation meets or exceeds that requirement.
Consequently, we define a key performance indicator of the time and user average flow gap ratio:
\begin{align}
    J_{f} =
    \dfrac{1}{T} \sum_{t\in\Tcal}
    \left(\dfrac{1}{N_{u}} \sum_{i\in\Vcalut} s_{i,t}\right).
    \label{eq:cost_satisfaction_flow}
\end{align}
It is important to note that this metric can be linearized by introducing a slack variable~\cite{Boyd_04}.

Let $\ell_{i}$ represent the latency of the $i$-th traffic and this is calculated as
\begin{align}
    \ell_{i,t} = \sum_{e\in\Ecalbar_{i}} \bar{\ell}_{e} y_{i,e,t},~\forall i\in\Vcalut,
    \label{eq:model_latency}
\end{align}
where $\bar{\ell}_{e}$ denotes the latency caused by propagation in link $e$.
This calculation includes the fixed latency between relay and destination gateways in terrestrial links.
Recall that, in this paper, ``latency'' refers solely to the propagation time.
To evaluate performance regarding latency, we use the time and user average latency:
\begin{align}
    J_{\ell} =
    \dfrac{1}{T} \sum_{t\in\Tcal}
    \left(\dfrac{1}{N_{u} L} \sum_{i\in\Vcalut} \ell_{i,t}\right),
    \label{eq:cost_latency}
\end{align}
where $L$ is a normalization constant for the latency and is chosen so that this value is between 0 and 1.

Therefore, based on Eqs.~in (\ref{eq:cost_gateway_number}) - (\ref{eq:cost_latency}), we define the cost function as:
\begin{align}
    J = w_{g} J_{g} + w_{f} J_{f} + w_{\ell} J_{\ell},
\label{eq:cost}
\end{align}
where $w_{g} \geq 0, w_{f} \geq 0$, and $w_{\ell} \geq 0$ are the weights for each term and chosen so that $w_{g} + w_{f} + w_{\ell} = 1$ holds.
Satellite operators determine these parameters based on their operational policies.

\subsection{Optimization Problem}

We formulate a joint gateway placement and routing optimization problem by the following:
\begin{align}
    \begin{aligned}
        & \underset{\substack{x_{\ell}, \{b_{i,t}, y_{i,e,t}, \\ f_{i,e,t}, z_{e,t}\}_{t\in\Tcal}}}{\text{minimize}}~~~
        && J~\text{in Eq.~(\ref{eq:cost})} \nn \\
        & \text{~~subject~to}
        && \text{Eqs.~(\ref{eq:constraint_feasibility_y_x}) - (\ref{eq:constraint_flow_gateway_others})}
    \end{aligned}
\end{align}

This problem is a mixed-integer linear problem and NP-hard because the decision variables include binary variables, and the cost function and constraints are linear in all the decision variables.
This type of problem can be efficiently solved using heuristic algorithms such as the branch-and-cut algorithm, which combines the branching and bounding steps of the branch-and-bound algorithm with cutting planes to prune the solution space effectively~\cite{Stubbs_MP99}.

\section{Numerical Simulation}

\begin{table}[b]
\centering
\caption{Simulation parameters.}
    \scalebox{1}{
    \begin{tabular}{ll} \hline 
    Parameters & Value \\ \hline \hline
    Simulation total time & 1800~s \\
    Simulation step & 60~s \\
    Total time duration $T$ & 31 \\
    Gateway candidate and user positions & Fig.~\ref{fig:world_map_gateway_user} \\
    Number of gateway candidates $N_{g}$ & 10 \\
    Number of users $N_{u}$ & 20 \\
    Flow requirement of users & 50~Mbps \\
    Number of satellites $N_{s}$ & 60 \\
    Number of planes $P$ & 6 \\
    Number of satellites in each plane $K$ & 10 \\
    Altitude & 800~km \\
    Inclination, Eccentricity, RAAN & 55, 0, 30~deg \\
    Capacity of user links $\ulcapacity$ & 250~Mbps \\
    Capacity of inter-satellite links $\islcapacity$ & 1~Gbps \\
    Capacity of feeder links $\flcapacity$ & 500~Mbps \\
    Normalization constant for the latency $L$ & 0.1~s \\
    \hline
    \end{tabular}
    }
\label{table:simulation_parameters}
\end{table}

This section presents the results of the proposed joint optimization for gateway placement and routing.
We derived available accesses of user, inter-satellite, and feeder links by using the Systems Tool Kit (STK). 
We solved the problem defined in the previous section by using the Gurobi Optimizer for Python, which implements the branch-and-cut algorithm.

\begin{figure}[tb]
    \centering
    \includegraphics[width=0.98\columnwidth]{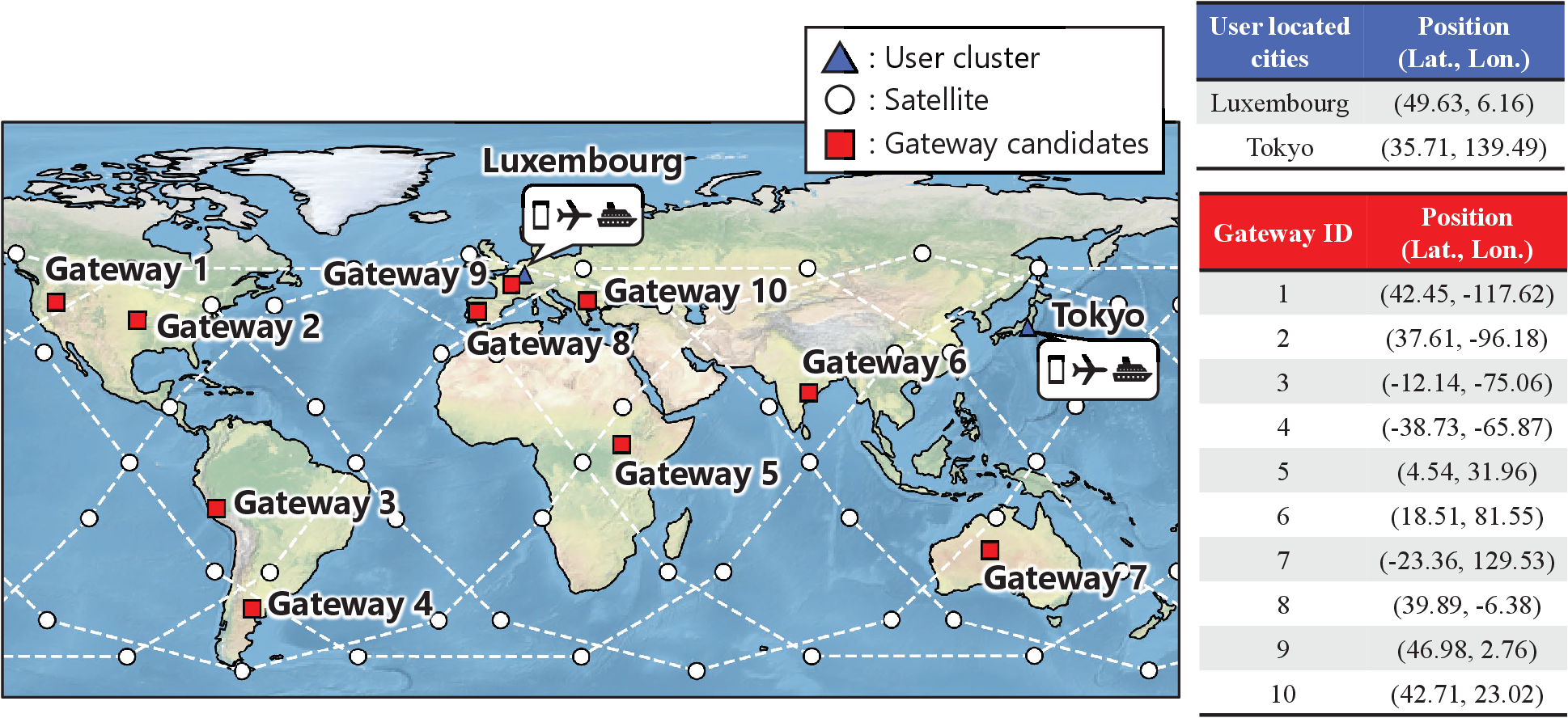}
    \caption{Placement of ten gateway candidates and two cities where the user clusters are located.}
\label{fig:world_map_gateway_user}
\end{figure}

The simulation parameters are detailed in Table~\ref{table:simulation_parameters}.
In the simulation, we set the parameters as $N_{u} = 20$, $N_{s} = 60$, and $N_{g} = 10$.
The satellites are equally spaced within each plane, with the argument of perigee and the true anomaly of each satellite in each plane set to be equally spaced accordingly.
We set ten of the 16 candidate gateways identified in \cite{Baeza_IEEENetwork23}, as depicted in Fig.~\ref{fig:world_map_gateway_user}.
A user cluster is located in Luxembourg at coordinates (49.63, 6.16) and in Tokyo, Japan at coordinates (35.71, 139.49), comprising ten users in each location.
Each user generates a constant traffic demand of $r_{i} = 50~\text{Mbps}$, and this value is time-invariant for the whole time duration.
The destination gateways are designated for each user located in these two cities, with ten gateways assigned to each group.

For evaluating the proposed optimization method, we compare the performance of three cases by varying the weights in the cost function as shown in Table~\ref{table:results_num_gateway}; 
$(w_{g}, w_{f}, w_{\ell}) = (0.5, 0.4, 0.1), (0.3, 0.4, 0.3),$ and $(0.1, 0.4, 0.5)$ for Cases~A, B, and C, respectively.
This configuration implies that Case~A prioritizes the number of gateways more, while Case~C emphasizes the traffic latency more, and Case~B is in the middle of them.
In this setting, we set the fixed value for the weight of the flow term $w_{f}$.

\begin{table}[tb]
\centering
\caption{Comparing results of the number of active gateways and the average latency in three cases.}
    \scalebox{0.95}{
    \begin{tabular}{cccc} \hline 
    \multirow{2}{*}{Case} & Weights & Number of active gateways & Time and user  \\
    & $(w_{g}, w_{f}, w_{\ell})$ & (Selected gateway ID) & average latency \\ \hline \hline
    A & (0.5, 0.4, 0.1) & 2 (1, 6) & 69.9~ms \\
    B & (0.3, 0.4, 0.3) & 3 (1, 5, 6) & 59.8~ms \\
    C & (0.1, 0.4, 0.5) & 5 (1, 3, 5, 6, 8) & 47.6~ms \\
    \hline
    \end{tabular}
    }
\label{table:results_num_gateway}
\end{table}

\begin{figure}[tb]
    \centering
    \includegraphics[width=0.98\columnwidth]{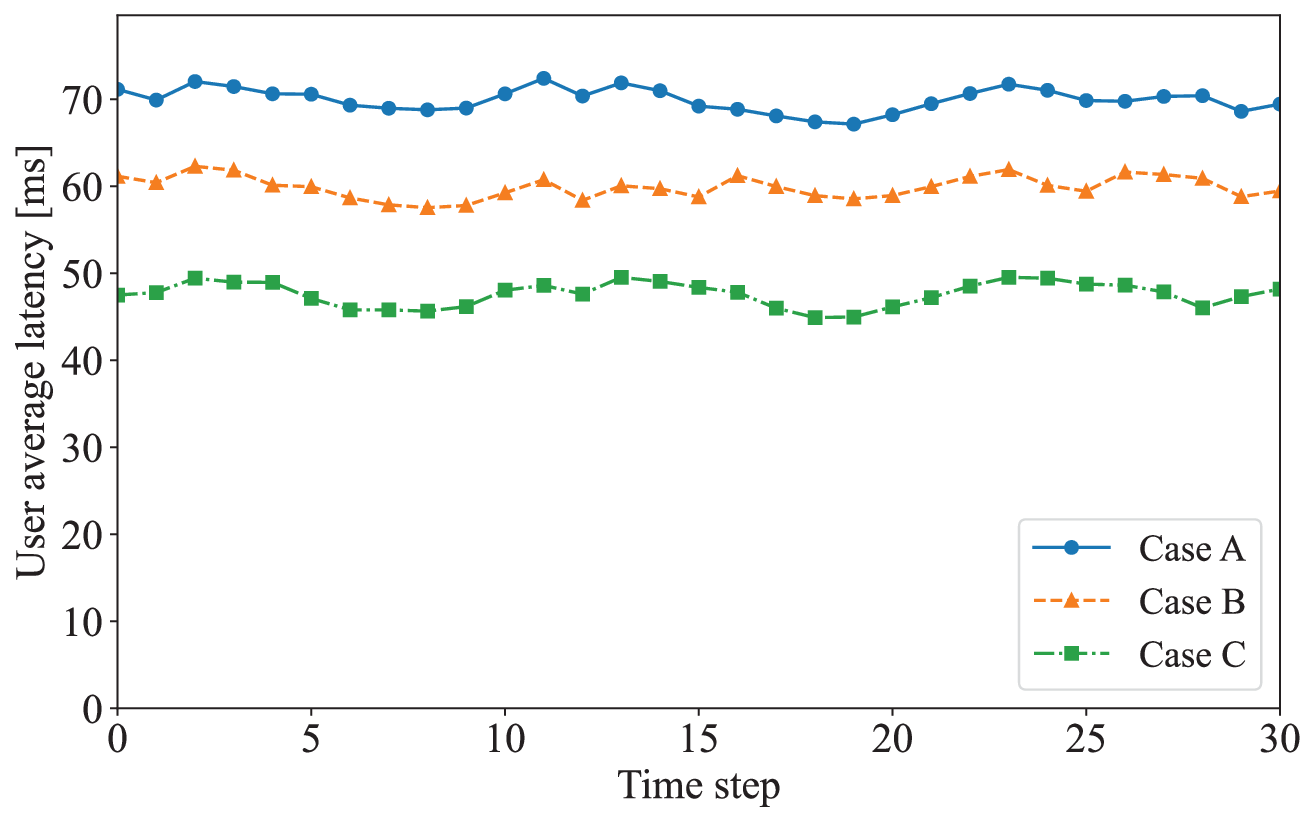}
    \caption{The results of the user average latency in three cases.}
\label{fig:results_average_latency}
\end{figure}

We show the comparing results of these three cases in Table~\ref{table:results_num_gateway} and the user average latency in Fig.~\ref{fig:results_average_latency}.
For Case~A, the number of active gateways is the least, and that number increased as we set less priority for the number of them.
On the other hand, in Case~C, the time and average latency of traffic decreased because we prioritized the traffic latency more compared to the other cases.
In this result, Gateways~1 and 6 were always selected because they were each geographically distributed enough to allow for moderate latency of the traffic they accommodated.
Thus, we can conclude these two were more dominant than the others in this setting.

\begin{figure}[tb]
    \centering
    \includegraphics[width=0.98\columnwidth]{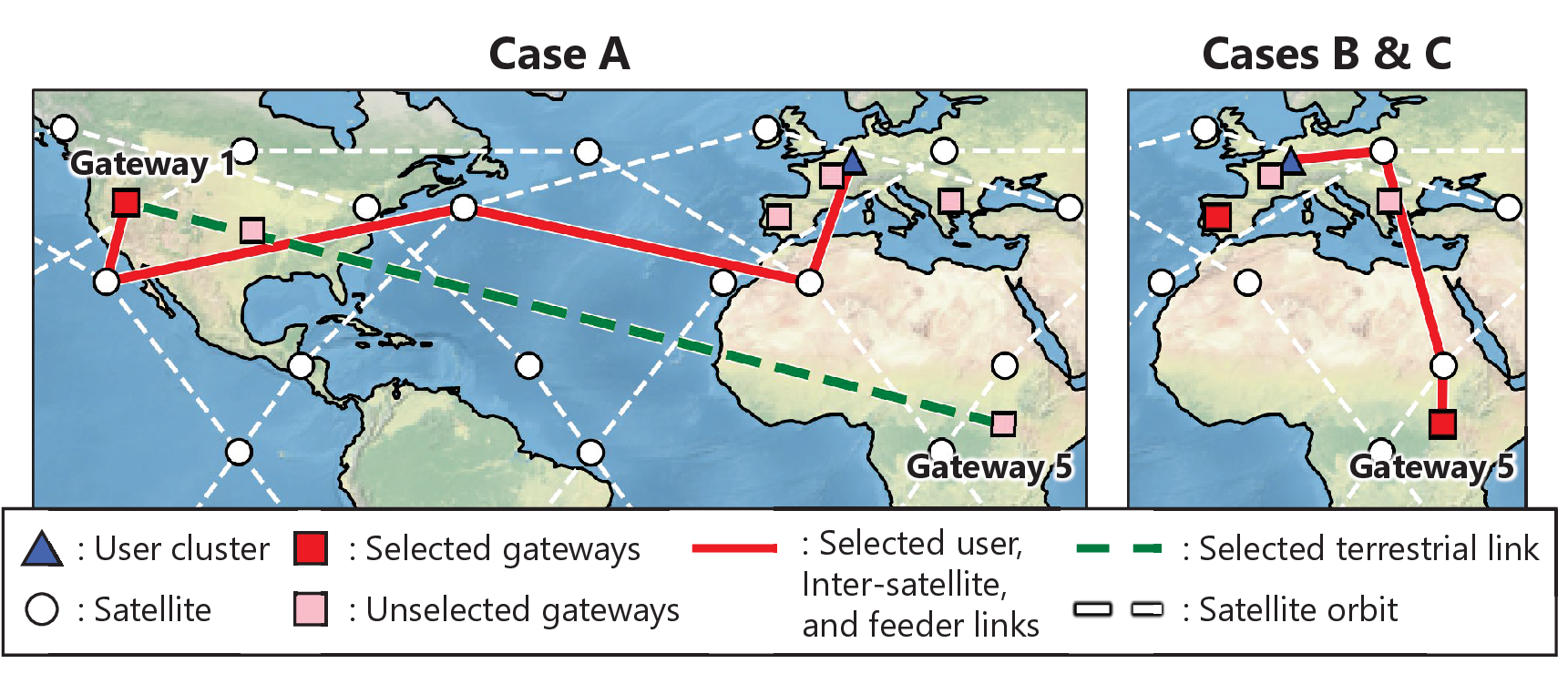}
    \caption{The optimized routing results of a certain user in Luxembourg in each case.}
\label{fig:results_routing}
\end{figure}

Fig.~\ref{fig:results_routing} shows the optimized routing results at a specific time step for one user in Luxembourg, whose destination gateway is Gateway~5 at the center of Africa.
For Cases~B and C, Gateway 5 was selected as an active one so the satellites could send the traffic to that gateway via a feeder link.
However, the traffic had a longer route in Case~A because Gateway~5 was not selected, and that traffic had to be routed via Gateway~1 in the U.S., which worked as a relay gateway, and experienced longer latency in terrestrial links.
From these results, we also conclude the link utilization efficiency is unbalanced, and the risk of traffic loss in case of failure is also higher when the number of selected gateways is small because feeder links are concentrated on fewer satellites.

\section{Conclusion}

In this paper, we proposed a joint optimization framework for the number of gateways and network performance.
We modeled a satellite and ground network model, incorporating users, satellites, and gateways, and defined the optimization problem with a cost function that combines the number of gateways, flow allocation, and traffic latency.
This approach enables satellite operators to set their operational priorities in alignment with their policies.
Our comparative analysis of different weights in the cost function verified the proposed method and demonstrated that it reduced the number of gateways.
In future work, we will explore the integration of machine learning algorithms to refine the optimization process further.

\section*{Acknowledgment}
Y. Abe acknowledges Mr.~Almoatssimbillah Saifaldawla from the University of Luxembourg for helpful advice on modeling a satellite constellation in MATLAB.
The work of F. Ortiz, E. Lagunas and S. Chatzinotas has been supported by the Luxembourg National Research Fund (FNR) under the project MegaLEO (C20/IS/14767486).

\bibliographystyle{IEEEtran}
\bibliography{myref}

\begin{thebibliography}{10}
\providecommand{\url}[1]{#1}
\csname url@samestyle\endcsname
\providecommand{\newblock}{\relax}
\providecommand{\bibinfo}[2]{#2}
\providecommand{\BIBentrySTDinterwordspacing}{\spaceskip=0pt\relax}
\providecommand{\BIBentryALTinterwordstretchfactor}{4}
\providecommand{\BIBentryALTinterwordspacing}{\spaceskip=\fontdimen2\font plus
\BIBentryALTinterwordstretchfactor\fontdimen3\font minus
  \fontdimen4\font\relax}
\providecommand{\BIBforeignlanguage}[2]{{%
\expandafter\ifx\csname l@#1\endcsname\relax
\typeout{** WARNING: IEEEtran.bst: No hyphenation pattern has been}%
\typeout{** loaded for the language `#1'. Using the pattern for}%
\typeout{** the default language instead.}%
\else
\language=\csname l@#1\endcsname
\fi
#2}}
\providecommand{\BIBdecl}{\relax}
\BIBdecl

\bibitem{Al-Hraishawi_CST23}
H.~Al-Hraishawi, H.~Chougrani, S.~Kisseleff, E.~Lagunas, and S.~Chatzinotas,
  ``A survey on nongeostationary satellite systems: The communication
  perspective,'' \emph{IEEE Communications Surveys \& Tutorials}, vol.~25,
  no.~1, pp. 101--132, 2023.

\bibitem{Guo_TVT21}
J.~Guo, D.~Rinc{\'{o}}n, S.~Sallent, L.~Yang, X.~Chen, and X.~Chen, ``Gateway
  placement optimization in {LEO} satellite networks based on traffic
  estimation,'' \emph{IEEE Transactions on Vehicular Technology}, vol.~70,
  no.~4, pp. 3860--3876, 2021.

\bibitem{Chen_IOTJ22}
Q.~Chen, L.~Yang, J.~Guo, X.~Liu, and X.~Chen, ``Optimal gateway placement for
  minimizing intersatellite link usage in {LEO} megaconstellation networks,''
  \emph{IEEE Internet of Things Journal}, vol.~9, no.~22, pp. 22\,682--22\,694,
  2022.

\bibitem{Portillo_IEEEAeroConf18}
I.~del Portillo, B.~Cameron, and E.~Crawley, ``Ground segment architectures for
  large {LEO} constellations with feeder links in ehf-bands,'' in
  \emph{Proceedings of the 2018 IEEE Aerospace Conference}, 2018.

\bibitem{Chen_IJSCN21}
Q.~Chen, L.~Yang, X.~Liu, J.~Guo, S.~Wu, and X.~Chen, ``Multiple gateway
  placement in large-scale constellation networks with inter-satellite links,''
  \emph{International Journal of Satellite Communications and Networking},
  vol.~39, no.~1, pp. 47--64, 2021.

\bibitem{Zhu_VTC20}
C.~Zhu, Y.~Li, M.~Zhang, Q.~Wang, and W.~Zhou, ``An optimization method for the
  gateway station deployment in {LEO} satellite systems,'' in \emph{Proceedings
  of the 2020 IEEE 91st Vehicular Technology Conference}, 2020.

\bibitem{Baeza_IEEENetwork23}
V.~M. Baeza, F.~Ortiz, E.~Lagunas, T.~S. Abdu, and S.~Chatzinotas, ``Gateway
  station geographical planning for emerging non-geostationary satellites
  constellations,'' \emph{IEEE Network}, Accepted for publication, DOI:
  10.1109/MNET.2023.3321531.

\bibitem{Bazaraa_09}
M.~S. Bazaraa, J.~J. Jarvis, and H.~D. Sherali, \emph{Linear Programming and
  Network Flows (4th Edition)}.\hskip 1em plus 0.5em minus 0.4em\relax Wiley,
  2009.

\bibitem{Boyd_04}
S.~Boyd and L.~Vandenberghe, \emph{Convex Optimization}.\hskip 1em plus 0.5em
  minus 0.4em\relax Cambridge University Press, 2004.

\bibitem{Stubbs_MP99}
R.~A. Stubbs and S.~Mehrotra, ``A branch-and-cut method for 0-1 mixed convex
  programming,'' \emph{Mathematical Programming}, vol.~86, pp. 515--532, 1999.

\end{thebibliography}

\end{document}